\documentclass[twocolumn,english]{revtex4-1}
\usepackage[T1]{fontenc}
\usepackage[latin9]{inputenc}
\usepackage{color}
\usepackage{units}
\usepackage{amsmath}
\usepackage{cancel}
\usepackage{graphicx}
\usepackage{esint}
\PassOptionsToPackage{normalem}{ulem}
\usepackage{ulem}

\makeatletter
\usepackage{amsmath}
\makeatother

\usepackage{babel}
\begin{document}

\title{Phenomenological theory of heat transport in the fractional quantum
Hall effect}

\author{Amit Aharon-Steinberg, Yuval Oreg and Ady Stern }

\address{Department of Condensed Matter Physics, Weizmann Institute of Science,
Rehovot 76100, Israel}
\begin{abstract}
The thermal Hall conductance is a universal and topological property
which characterizes the fractional quantum Hall (FQH) state. The quantized
value of the thermal Hall conductance has only recently been measured
experimentally in integer quantum Hall (IQH) and FQH regimes, however,
the existing setup is not able to detect if the thermal current is
counter-propagating or co-propagating with the charge current. Furthermore,
although there is experimental evidence for heat transfer between
the edge modes and the bulk, the current theories do not take this
dissipation effect in consideration. In this work we construct phenomenological
rate equations for the heat currents which include equilibration processes
between the edge modes and energy dissipation to an external thermal
bath. Solving these equations in the limit where temperature bias
is small, we compute the temperature profiles of the edge modes in
a FQH state, from which we infer the two terminal thermal conductance
of the state as a function of the coupling to the external bath. We
show that the two terminal thermal conductance depends on the coupling
strength, and can be non-universal when this coupling is very strong.
Furthermore, we propose an experimental setup to examine this theory,
which may also allow the determination of the sign of the thermal
Hall conductance.
\end{abstract}
\maketitle

\subsection{Introduction}

The fractional quantum Hall (FQH) state is a topological state of
matter, and therefore it is described by universal and topological
properties~\cite{Prange,Wen}. Two such properties are the Hall conductance
and the thermal Hall conductance~\cite{Kane1995}. In Abelian FQH
states, which are described by an integer valued symmetric matrix,
termed the $K$ matrix, these two topological properties relate to
the edge modes and the $K$ matrix in different ways. The Hall conductance,
given by $\sigma_{H}=\nu\frac{e^{2}}{h}$, where $\nu$ is the filling
fraction, is governed by the downstream charge mode~\cite{Wen,Kane1995,Stern2005,Kane1994}.
However, the thermal Hall conductance, given by 
\begin{equation}
\kappa_{H}=n_{\text{net}}\kappa_{0}T,\label{eq:K_xy}
\end{equation}
where $\kappa_{0}=\frac{\pi^{2}k_{B}^{2}}{3h}$ is the quantum of
thermal conductance~\cite{Pendry1983} and $T$ is the temperature,
is governed by the net number of edge modes, $n_{\text{net}}=n_{d}-n_{u}$,
which is the difference between the number of downstream and upstream
modes in the edge theory~\cite{Kane1997}.

An interesting phenomenon occurs in hole-like states, which have $\frac{1}{2}<\nu<1$.
Theory suggests that these states are characterized by $n_{u}\geq n_{d}$.
In the case of $n_{u}>n_{d}$, when the modes are at equilibrium,
charge and heat flow on different edges of the FQH liquid. Whereas,
in the case of $n_{u}=n_{d}$, the heat flow is diffusive~\cite{Kane1997,Nosiglia2018,Protopopov}.

The thermal Hall conductance has only recently
been measured in IQH states~\cite{Jezouin2013,Banerjee2017}, in
FQH states~\cite{Banerjee2017,Banerjee2018} and in the magnetic
material $\text{\ensuremath{\alpha}-RuC\ensuremath{l_{3}}}$~\cite{Kasahara2018}.
However, the current setups used in these experiments cannot determine
which edge carries the heat current. Hence, the thermal Hall conductance
still holds more information about the $K$ matrix, which was yet
realized. Furthermore, it was shown experimentally that there is energy
dissipation from the edge modes of a QH state~\cite{Venkatachalam2012a}.
Energy dissipation can arise from different mechanisms.
Electron-electron interaction, for example, which is accountable for
the appearance of the charge and neutral modes~\cite{Meir1994,LeSueur2010,Altimiras2010,Dolev2011,Bid2011,Gross2012,Gurman2012,Wang2013,Grivnin2014,Inoue2014},
may cause inter-edge-modes energy relaxation~\cite{LeSueur2010,Altimiras2010},
but may also account for energy loss to puddles in the bulk. Electron-phonon
interaction may lead to energy dissipation from the edge modes~\cite{LeSueur2010,Altimiras2012,Venkatachalam2012a}.
Nonetheless the present theories~\cite{Kane1997,Simon2018,Feldman2018,Mross}
neglect such contribution to the heat transport of the QH state. Such
energy dissipation, which may alter the thermal Hall conductance of
the state, should therefore be incorporated into the theory. 

In this manuscript we develop a phenomenological theory for the heat
transport in the edge modes of a FQH state, which elaborates on the
phenomenological equations derived in Ref.~\cite{Banerjee2017},
and on the theoretical analysis performed in Refs.~\cite{Nosiglia2018,Protopopov},
in order to include dissipation from the edge modes to an external
thermal bath. We note that recent theoretical analysis of the observation of quantized thermal Hall conductance in the magnetic material $\text{\ensuremath{\alpha}-RuC\ensuremath{l_{3}}}$~\cite{Vinkler-Aviv2018,Balents2018}, take into account coupling to phonons. However, in this three dimensional system, and in the experimental set-up employed in Ref.~\cite{Kasahara2018}, energy transfered to the phonons is not lost, and is included in the measured heat current. In this manuscript, and in the experiment carried out in Refs.~\cite{Banerjee2017,Banerjee2018} energy transfered to phonons, or any other mode of dissipation, leaks out of the system and is not measured. By solving the heat transport rate equations for small
temperature difference between the two sides of the FQH liquid, we
find the temperature profiles of the edge modes, as a function of
the equilibration length, $\xi_{e}$, between the modes, the dissipation
length, $\xi_{d}$, for energy dissipation to the external bath, and
the system size, $L$. We then define and calculate the two terminal
thermal conductance, and show that its measurement may strongly depend
on the dissipation length. Since we are interested in exploring the
effect of dissipation on the topological thermal Hall conductance,
it is required that $L\gg\xi_{e}$. While $L$
usually varies between tens to hundreds of $\mu m$, it was found
experimentally that typical $\xi_{e}$ can vary between $3\mu m$~\cite{LeSueur2010}
and $20\mu m$~\cite{Banerjee2017}, depending
on the temperature, and that $\xi_{d}$ is bounded from below by $30\mu m$~\cite{LeSueur2010}.
We find that for $\xi_{d}\gg\xi_{e}$ the two
terminal conductance approaches the topological and universal value
of Eq.~\eqref{eq:K_xy}, whereas for $\xi_{d}\ll\xi_{e}$ the two
terminal thermal conductance is not universal anymore, and is sensitive
to edge reconstruction processes.

Furthermore, we propose an experimental setup to test this phenomenological
theory, and to determine the sign of the thermal Hall conductance.
This experimental setup relies on quantum dots (QDs) which are coupled
to the edges of a FQH state. By exploiting a relation between the
thermoelectric coefficient and the electric conductance of the QDs~\cite{Roura-Bas2017},
we point out that they can be used as local thermometers for electrons
on the FQH edge state. The local temperatures of the edge states can
be deduced from a measurement of the thermoelectric current through
the QDs, and thus the temperature profiles can be measured. We show
that the sign of the thermal Hall conductance may be determined from
the measured temperature profiles.

\subsection{Phenomenological heat transport theory in Abelian FQH states}

\begin{figure}
\centering{}\includegraphics[bb=170bp 230bp 630bp 390bp,clip,scale=0.5]{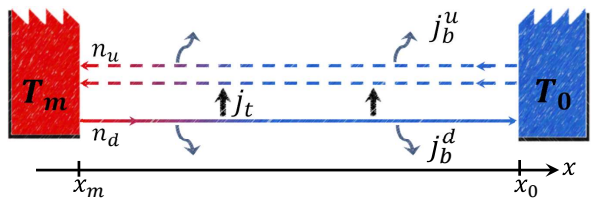}\caption{\textbf{\label{fig:two terminal schematic}}An illustration of the
lower edge of a FQH liquid in a two terminal system, with $n_{d}$
downstream modes (solid lines) and $n_{u}$ upstream mode (dashed
lines), propagating in the opposite direction. Temperature difference
is applied between the right and left contacts, $\Delta T=T_{m}-T_{0}>0$.
The black vertical arrows represent the equilibration current density
between the edge modes on the same edge. The blue wiggly arrows represent
the dissipation current density, to the external thermal bath. }
\end{figure}

Without loss of generality, we assume the directions of flow of the
edge modes on the lower edge of a FQH state are as depicted in Fig.
\eqref{fig:two terminal schematic}. On the upper edge $n_{u}\leftrightarrow n_{d}$
and the directions of flow of the edge modes are reversed. In this
situation the FQH state has $n_{d}$ downstream modes and $n_{u}$
upstream modes, which for this analysis are assumed to obey $n_{d}\neq n_{u}$.
We shall consider the case $n_{u}=n_{d}=1$ when we discuss the $\nu=\frac{2}{3}$
FQH state. The downstream modes on the lower edge are emanating from
an Ohmic contact at position $x_{m}$ at temperature $T_{m}$, and
the upstream modes are emanating from another Ohmic contact at position
$x_{0}$ at temperature $T_{0}$. Both Ohmic contacts are at the same
chemical potential. For $T_{m}>T_{0}$, the downstream modes are expected
to be hotter than the upstream modes. Thus energy will be transferred
from the downstream to the upstream modes, through a heat current
density $j_{t}$, in order to achieve equilibration. In addition,
in order to model dissipation, we assume the edge modes are coupled
to an external thermal bath kept at temperature $T_{0}$, such that
there are dissipation current densities $j_{b}^{d}$ and $j_{b}^{u}$
from the downstream and upstream modes to the bath. Assuming that
energy is conserved in the system composed of the edge modes and the
external bath, the heat currents flowing through the 1D downstream
modes and the 1D upstream modes, denoted by $J_{d}$ and $J_{u}$
respectively, are described by the following rate equations:

\begin{equation}
\begin{aligned}J_{d}\left(x+\delta x\right) & =J_{d}\left(x\right)-j_{t}\left(x\right)\delta x-j_{b}^{d}\left(x\right)\delta x\\
J_{u}\left(x\right) & =J_{u}\left(x+\delta x\right)+j_{t}\left(x\right)\delta x-j_{b}^{u}\left(x\right)\delta x.
\end{aligned}
\label{eq:CurrentsEq}
\end{equation}

\subsubsection*{Temperature profiles}

The temperature dependencies of the heat currents in Eq.~\eqref{eq:CurrentsEq}
are modeled as follows. The heat current flowing in the 1D downstream
and upstream edge modes is modeled as $J_{i}(x)=\frac{1}{2}\kappa_{0}n_{i}T_{i}^{2}(x)$~\cite{Kane1997},
where $i=d,u$. The equilibration current density
$j_{t}$ is modeled by Newton's law of cooling, $j_{t}\left(x\right)=\frac{1}{2}\frac{\kappa_{0}}{\xi_{e}}\left[T_{d}^{2}\left(x\right)-T_{u}^{2}\left(x\right)\right]$,
where $\xi_{e}$ is the relaxation length, similarly to Ref.~\cite{Banerjee2017}.
The dissipation current to the external thermal bath is modeled by
a temperature power law relative to the bath temperature: $j_{b}^{i}(x)=\frac{1}{2}\kappa_{0}n_{i}B(T_{i}^{\alpha}(x)-T_{0}^{\alpha})$.
The exponent $\alpha$ has different values depending on the mechanism
of dissipation. Energy transfer from electron to phonons, for example,
may lead to $\alpha=5$, but also to smaller values depending on details~\cite{Wellstood1994}.
Electron-electron interaction gives $1<\alpha<2$, depending on the
extent to which impurities are involved~\cite{Imry}. To simplify
the solution and further treatment, we write the equations using the
dimensionless parameter: $\tau_{i}\left(x\right)=\frac{T_{i}^{2}\left(x\right)}{T_{0}^{2}}$,
and we denote: $\beta=BT_{0}^{\alpha-2}$. Then, the equations can
be written as a set of coupled differential equations for $\tau_{u}$
and $\tau_{d}$:
\begin{equation}
\begin{aligned}\frac{d\tau_{d}}{dx} & =-\frac{1}{n_{d}\xi_{e}}\left(\tau_{d}\left(x\right)-\tau_{u}\left(x\right)\right)-\beta\left(\tau_{d}^{\frac{\alpha}{2}}\left(x\right)-1\right)\\
\frac{d\tau_{u}}{dx} & =-\frac{1}{n_{u}\xi_{e}}\left(\tau_{d}\left(x\right)-\tau_{u}\left(x\right)\right)+\beta\left(\tau_{u}^{\frac{\alpha}{2}}\left(x\right)-1\right).
\end{aligned}
\label{eq:GeneralEq}
\end{equation}
The temperature dependence of the heat currents to the thermal bath
and the exchange current are expected to hold for small temperature
difference, $\Delta T=T_{m}-T_{0}$. The boundary conditions are:
\begin{eqnarray}
\tau_{d}\left(x_{m}\right)=\tau_{m}=\frac{T_{m}^{2}}{T_{0}^{2}} & \quad;\quad & \tau_{u}\left(x_{0}\right)=1=\frac{T_{0}^{2}}{T_{0}^{2}}.\label{eq:BoundaryC}
\end{eqnarray}

An analytic solution to Eqs.~\eqref{eq:GeneralEq}, with the boundary
conditions given by Eq.~\eqref{eq:BoundaryC} can be obtained for
small temperature difference from $T_{0}$, i.e. $\Delta T\ll T_{0}$,
such that $\tau_{i}\left(x\right)=1+\delta\tau_{i}\left(x\right)$.
Linearizing the equations, we find a new interaction parameter, $\frac{1}{\xi_{d}}=\frac{\beta\alpha}{2}$,
which we call the dissipation length. Integrating the linearized differential
equations with the appropriate boundary conditions, $\tau_{d}\left(x\right)$
and $\tau_{u}\left(x\right)$ of the lower edge are obtained: 

\begin{widetext}\begin{subequations}

\begin{align}
\tau_{d}^{\text{lower}}\left(x\right) & =1+\frac{\left(\frac{N}{2\bar{n}}+\frac{\xi_{e}}{\xi_{d}}\right)\sinh\left[\Lambda\left(x_{0}-x\right)\right]+\Lambda\xi_{e}\cosh\left[\Lambda\left(x_{0}-x\right)\right]}{\left(\frac{N}{2\bar{n}}+\frac{\xi_{e}}{\xi_{d}}\right)\sinh\left[\Lambda L\right]+\Lambda\xi_{e}\cosh\left[\Lambda L\right]}e^{-\frac{x-x_{m}}{2\bar{n}\xi_{e}}}\left(\tau_{m}-1\right)\label{eq:TemprofileTd}\\
\tau_{u}^{\text{lower}}\left(x\right) & =1+\frac{1}{n_{u}}\frac{\sinh\left[\Lambda\left(x_{0}-x\right)\right]}{\left(\frac{N}{2\bar{n}}+\frac{\xi_{e}}{\xi_{d}}\right)\sinh\left[\Lambda L\right]+\Lambda\xi_{e}\cosh\left[\Lambda L\right]}e^{-\frac{x-x_{m}}{2\bar{n}\xi_{e}}}\left(\tau_{m}-1\right),\label{eq:TemprofileTu}
\end{align}
\end{subequations}\end{widetext} where $L=x_{0}-x_{m}$, $\bar{n}=\frac{n_{u}n_{d}}{n_{u}-n_{d}}$,
$N=\frac{n_{u}+n_{d}}{n_{u}-n_{d}}$ and $\Lambda=\frac{1}{2\bar{n}\xi_{e}}\sqrt{1+4\bar{n}^{2}\frac{\xi_{e}}{\xi_{d}}\left(\frac{N}{\bar{n}}+\frac{\xi_{e}}{\xi_{d}}\right)}$.
 To determine $\tau_{d/u}\left(x\right)$ on the upper edge, the
number of edge modes needs to be interchanged, $n_{u}\leftrightarrow n_{d}$,
and for consistency with the direction of chirality also $\tau_{d}\leftrightarrow\tau_{u}$,
such that $\tau_{d}^{\text{lower}}\left(x;n_{d},n_{u}\right)=\tau_{u}^{\text{upper}}\left(x;n_{u},n_{d}\right)$.
Numerically we can go beyond the linearized regime, however in doing
so we found that small deviations from that regime do not change the
qualitative picture.

\subsubsection*{Normalized two terminal thermal conductance}

Assuming that heat can be transported from the hot contact to the
system only through the edge modes, the normalized two terminal thermal
conductance, $\kappa$, is defined according to:
\begin{equation}
J_{Q}=\frac{1}{2}\kappa_{0}\kappa\left(T_{m}^{2}-T_{0}^{2}\right),
\end{equation}
where $J_{Q}$ is the total heat current emanating from the hot contact
to the system, due to $\Delta T$. This $\kappa$ is composed of two
parts, corresponding to the heat flowing along the upper and lower
edges, which by assumption do not interact. Due to energy conservation,
the sum of the heat flowing in the edge modes and the integrated heat
dissipated to the thermal bath should not depend on the position along
the edge. Therefore, the contribution of the lower edge to the two
terminal thermal conductance is:
\begin{equation}
\kappa_{\text{lower}}=\frac{J_{d}\left(x\right)-J_{u}\left(x\right)-J_{p}+\intop_{x_{m}}^{x}\left[j_{b}^{d}\left(x'\right)+j_{b}^{u}\left(x'\right)\right]dx'}{\frac{1}{2}\kappa_{0}\left(T_{m}^{2}-T_{0}^{2}\right)},\label{eq:general_K}
\end{equation}
where $J_{p}=\frac{1}{2}\kappa_{0}\left(n_{d}-n_{u}\right)T_{0}^{2}$
is the persistent heat current in the system at equilibrium, which
has no divergence because the upper edge has an opposite term. It
is subtracted from both edges in order to expose the net current above
the equilibrium current flowing in the system due to the chirality.

The normalized two terminal thermal conductance of the system is obtained
by summing the contributions from both edges. Plugging the temperature
dependencies, given by Eqs.~\eqref{eq:TemprofileTd} and~\eqref{eq:TemprofileTu},
the two terminal thermal conductance is readily obtained:

\begin{widetext}
\begin{equation}
\kappa\left(\xi_{d},\xi_{e},L\right)=\kappa_{\text{lower}}+\kappa_{\text{upper}}=\frac{1}{2\bar{n}}\frac{n_{u}e^{\Lambda L}+n_{d}e^{-\Lambda L}}{\Lambda\xi_{e}\cosh\left(\Lambda L\right)+\left(\frac{N}{2\bar{n}}+\frac{\xi_{e}}{\xi_{d}}\right)\sinh\left(\Lambda L\right)}+\left(n_{u}+n_{d}\right)\frac{\left(\Lambda\xi_{e}-\frac{1}{2\bar{n}}\right)\cosh\left(\Lambda L\right)+\frac{\xi_{e}}{\xi_{d}}\sinh\left(\Lambda L\right)}{\Lambda\xi_{e}\cosh\left(\Lambda L\right)+\left(\frac{N}{2\bar{n}}+\frac{\xi_{e}}{\xi_{d}}\right)\sinh\left(\Lambda L\right)}.\label{eq:2termThermalCond}
\end{equation}
\end{widetext}

There are three competing length scales in our problem: the system
size $L$, the equilibration length $\xi_{e}$, and the dissipation
length $\xi_{d}$. Since we wish to discuss the thermal Hall conductance,
defined for a fully equilibrated edge system, it is required that
$L\gg\xi_{e}$, so that the edge modes are able to equilibrate over
the length of the system. Let us now elaborate more on the temperature
profiles and $\kappa$ of the hole like states, for both cases: (i)
$n_{u}>n_{d}$ and (ii) $n_{d}=n_{u}=1$ (corresponding to the $\nu=\frac{2}{3}$
state). 

\begin{figure*}
\centering{}\includegraphics[bb=0bp 120bp 792bp 470bp,clip,scale=0.55]{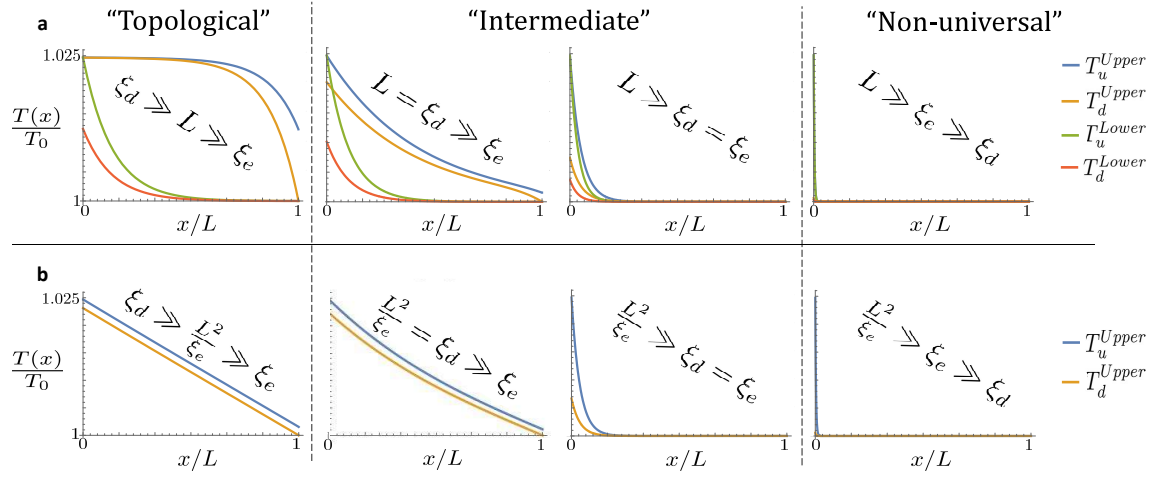}\caption{\textbf{\label{fig:Temperature-profiles}a. }Temperature profiles
of the downstream and upstream modes on the upper and lower edges
of a $\nu=\frac{3}{5}$ FQH state, described by $n_{d}=1$ and $n_{u}=2$.
The temperature profiles are plotted for different values of $\xi_{d}$
relative to $L$ and $\xi_{e}$, where $L=300\mu m$ and $\xi_{e}=20\mu m$~\cite{Banerjee2017}.
\textbf{b.} Temperature profiles of the downstream and upstream modes
of a $\nu=\frac{2}{3}$ FQH state, described by $n_{d}=n_{u}=1$.
The upper and lower edges exhibit the same temperature profile. }
\end{figure*}

\paragraph{Hole-like states with $n_{u}>n_{d}$ - \protect \\
}

The temperature profiles of the edge modes are given by Eqs~\eqref{eq:TemprofileTd}
and~\eqref{eq:TemprofileTu}, and $\kappa$ is given by Eq.~\ref{eq:2termThermalCond}.
To illuminate the physics let us discuss the temperature profiles
{[}Fig. (2.a){]} and $\kappa$ in the following regimes:

\uline{Topological regime ($\xi_{d}\gg L\gg\xi_{e}$) -} The edge
modes exchange energy with one another, and equilibrate to the temperature
of the upstream modes. In this regime their dissipation of energy
to the thermal bath is small. The normalized two terminal thermal
conductance acquires the absolute value of the topological value~\cite{Kane1997}
with two corrections to leading orders: $\kappa=\left(n_{u}-n_{d}\right)\left[1+2\frac{n_{d}}{n_{u}}e^{-\frac{L}{\bar{n}\xi_{e}}}\right]+4\bar{n}n_{d}\frac{\xi_{e}}{\xi_{d}}$.
The first exponential correction is due to the finite system size
$L$, and the second algebraic correction is due to dissipation to
the bath, that happens all along the edge. 

\uline{Intermediate regime ($L\gg\xi_{d}\gg\xi_{e}$) -} Most energy
is dissipated to the thermal bath before arrival to the cold contact,
therefore the temperature profiles decrease to $T_{0}$ on both edges.
However, the edge modes exchange energy before dissipating it all
to the thermal bath. Thus, to leading order, $\kappa$ acquires the
absolute value of the topological value, with an algebraic correction
due to dissipation: $\kappa=\left(n_{u}-n_{d}\right)+4\bar{n}n_{d}\frac{\xi_{e}}{\xi_{d}}$.
This correction can be of the order of $\left(n_{u}-n_{d}\right)$,
so $\kappa$ is not universal in this case.

\uline{Non-universal regime ($L\gg\xi_{e}\gg\xi_{d}$) -} The edge
modes dissipate all their energy to the thermal bath and therefore
the temperature profiles decrease to $T_{0}$ very close to the hot
contact. The thermal conductance, $\kappa$, in this case is the total
number of edge modes leaving the hot contact, $n=n_{u}+n_{d}$, with
a correction due to a competition between $\xi_{e}$ and $\xi_{d}$:
$\kappa=\left(n_{u}+n_{d}\right)-\frac{\xi_{d}}{\xi_{e}}$. This happens
because the modes emanating from the hot contact on both edges dissipate
all the energy to the external thermal bath, thus the heat conductance
is limited by the total number of modes emanating the hot contact.
The number $n=n_{u}+n_{d}$ is not universal, due to processes such
as edge reconstruction \cite{Wan2003,Sabo2017}. This limit and the
limit of a very short system, i.e. $L\ll\xi_{e}$, are qualitatively
similar.

\paragraph{$\nu=\frac{2}{3}$ state - \protect \\
}

The temperature profiles of the edge modes of the $\nu=\frac{2}{3}$
state are obtained by taking the limit of $n_{u}\rightarrow n_{d}=1$
in Eqs.~\eqref{eq:TemprofileTd},~\eqref{eq:TemprofileTu}. Substituting
the temperature profiles into Eq. \ref{eq:general_K} we obtain $\kappa$
for the $\nu=\frac{2}{3}$ state:
\begin{equation}
\kappa=2\left[1-\frac{1}{1+\frac{\xi_{e}}{\xi_{d}}+\Lambda_{\frac{2}{3}}\xi_{e}\coth\left(\Lambda_{\frac{2}{3}}L\right)}\right],
\end{equation}
where $\Lambda_{\frac{2}{3}}=\frac{1}{\xi_{e}}\sqrt{\frac{\xi_{e}}{\xi_{d}}\left(2+\frac{\xi_{e}}{\xi_{d}}\right)}$.
 To illuminate the physics, let us discuss the temperature profiles
of the edge modes {[}Fig. (2.b){]} and $\kappa$ in the corresponding
three regimes: 

\uline{The topological regime ($\xi_{d}\gg\nicefrac{L^{2}}{\xi_{e}}\gg\xi_{e}$)
-} The system is diffusive, therefore the temperature profiles are
linear along the edges, with a constant difference. The thermal conductance,
$\kappa$, approaches the absolute value of the topological value~\cite{Kane1997}
with a leading order algebraic correction, due to a competition between
the equilibration length and the finite system size: $\kappa=\frac{2}{1+\frac{L}{\xi_{e}}}$.

\uline{The intermediate regime ($\nicefrac{L^{2}}{\xi_{e}}\gg\xi_{d}\gg\xi_{e}$)
-} The system dissipate energy to the thermal bath, therefore the
temperature profiles are exponential, rather than linear. The thermal
conductance, $\kappa$, approaches the absolute value of the topological
value, with a leading order algebraic correction, due to the competition
between the equilibration  and dissipation lengths: $\kappa=\frac{2}{1+\sqrt{2\frac{\xi_{d}}{\xi_{e}}}}$.

\uline{The non-universal regime ($\nicefrac{L^{2}}{\xi_{e}}\gg\xi_{e}\gg\xi_{d}$)
-} The edge modes dissipate all their energy to the thermal bath,
so the temperature profiles decrease to $T_{0}$ very close to the
hot contact. The thermal conductance, $\kappa$, approaches the non-universal
value of the total number of modes, with an algebraic correction due
to the competition between equilibration and dissipation: $\kappa=2-\frac{\xi_{d}}{\xi_{e}}$.

\subsection{Proposed experimental setup}

\begin{figure}
\centering{}\includegraphics[bb=220bp 180bp 565bp 450bp,clip,scale=0.45]{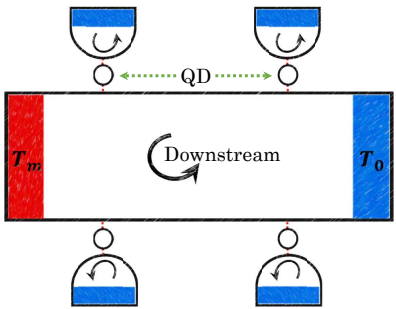}\caption{\label{fig:schematic pic}A schematic picture of the proposed experimental
setup. Two Ohmic contacts are connected to a Hall bar at a FQH state,
with a chirality defined by the solid black arrow. A temperature difference
$\Delta T=T_{m}-T_{0}$ is imposed between the contacts (Ref.~\cite{Banerjee2017,Jezouin2013}).
Multiple number of QDs, depicted by the full circles, are coupled
to both edges of the Hall bar. They are connected to Ohmic contacts,
thus enabling measurement of the thermoelectric currents that passes
through them.}
\end{figure}

This phenomenological theory may be tested by employing quantum dots
(QDs) as thermometers~\cite{Hoffmann2007,Venkatachalam2012a,Viola2012,Maradan2014}
for the temperature at various points along the edge. The proposed
experimental setup, depicted in Fig. \eqref{fig:schematic pic}, couples
QDs to the edges of FQH liquids, and is based on measuring the resulting
thermoelectric current. To deduce the temperature profiles from the
thermoelectric current, the thermoelectric coefficient needs to be
known. Following Furusaki~\cite{Furusaki1998}, the thermoelectric
coefficient, $G_{T},$ of a QD in a normal state, weakly coupled to
two FQH liquids, can be calculated to linear order in the temperature
difference between the two FQH states. In this regime, the thermoelectric
coefficient is found to be related to the conductance of the QD~\cite{Roura-Bas2017}
as 
\begin{equation}
G_{T}=\frac{\epsilon}{e}G,
\end{equation}
where $G$ is the linear electric conductance of the QD, $e$ is the
electron charge and $\epsilon$ is the energy difference between the
many body ground state energies of $N+1$ electrons and $N$ electrons
in the QD. Using this relation, the thermoelectric coefficient of
the QDs can be measured without applying a temperature bias. Thus
the temperature profiles can be deduced from the thermoelectric current
through the QDs, upon introduction of temperature difference $\Delta T$.

A measurement of the temperature profiles allows for the extraction
of the dissipation length, the equilibration length and the sign of
the thermal Hall conductance. For extraction of the latter, the system
needs to be in the topological regime ($\xi_{d}\gg L\gg\xi_{e}$).
In this regime, the edges are distinguished by their temperature profiles,
such that the edge which is expected to carry the heat current, according
to Ref.~\cite{Kane1997} is hotter {[}Fig. \eqref{fig:Temperature-profiles}{]}.

\subsection{Conclusions}

To conclude, the thermal Hall conductance is predicted to be a universal
and topological property of a FQH state, and therefore can help determining
the states in a more accurate way. Recent experiment has managed to
measure the absolute value of the thermal Hall conductance of Abelian
FQH states~\cite{Banerjee2017}, and consisted with the prediction
of Kane and Fisher~\cite{Kane1997} regarding these states. It should
be noted, however, that Ref~\cite{Kane1997} assumes the edge is
a closed system with respect to energy, while it was shown experimentally
that there can be energy dissipation from the edge \cite{Venkatachalam2012a}.

In this paper we elaborated on the phenomenological picture of the
temperature profiles of the edge modes of a FQH state with $n_{d}$
downstream mode and $n_{u}$ upstream modes described in Ref.~\cite{Banerjee2017},
by writing rate equations for heat transport through the edges, including
a dissipation term to an external thermal bath. By solving the phenomenological
equations, we found that the two terminal thermal conductance depends
on the coupling strength to the external thermal bath, in such a way
that when the coupling is extremely weak, the two terminal thermal
conductance acquires the universal topological value, however, when
the coupling is very strong the two terminal thermal conductance is
not universal anymore, and is subject to the influence of edge reconstruction
effects~\cite{Wan2003,Sabo2017}. 

Furthermore, we proposed to use QDs coupled to the edges of a FQH
state  to, first, test the above theory and measure the dissipation
length and the equilibration length, and second, to determine the
sign of the thermal Hall conductance. 
\begin{acknowledgments}
We would like to thank Mitali Banerjee, Dima E. Feldman, Moty Heiblum,
Tobias Holder, Gilad Margalit, David F. Mross, Amir Rosenblatt, Steven
H. Simon, Kyrylo Snizhko and Vladimir Umansky for constructive discussions.
We acknowledge support of the Israel Science Foundation; the European
Research Council under the European Communitys Seventh Framework Program
(FP7/2007-2013)/ERC Project MUNATOP; the DFG (CRC/Transregio 183,
EI 519/7-1). YO acknowledges support of the Binational Science Foundation.
AS acknowledges support of Microsoft Station Q.
\end{acknowledgments}

\bibliographystyle{apsrev4-1}
\bibliography{../pheno_theory_FQHE}

\end{document}